\documentclass[aps,prd,showpacs,twocolumn,floatfix]{revtex4}
\usepackage{graphicx}
\usepackage{dcolumn}
\usepackage{bm}
\begin{document}

\def\n{\noindent}
\def\be{\begin{equation}}    \def\ee{\end{equation}}
\def\bd{\begin{displaymath}} \def\ed{\end{displaymath}}
\def\ba{\begin{eqnarray}}    \def\ea{\end{eqnarray}}
\def\J{\rm J} 
\def\P{\rm P} 
\def\C{\rm C}
\def\m{m_p} 
\def\pp{p\bar p} 
\def\ppm{p\bar p m} 
\def\pppi{p\bar p \pi^0}
\def\ppf0{p\bar p f_0}
\def\ppeta{p\bar p \eta} 
\def\ppw{p\bar p \omega} 
\def\pprho0{p\bar p \rho^0} 
\def\ppphi{p\bar p \phi} 
\def\NN{N\bar N}
\def\NNm{N\bar N m} 
\def\NNV{N\bar N V} 
\def\nnw{NN\omega} 
\def\gNNpi{g_{NN\pi}}
\def\gw{g_{\omega}}
\def\aw{\alpha_{\omega}}
\def\kw{\kappa_{\omega}}
\def\gr{g_{\rho}}
\def\ar{\alpha_{\rho}}
\def\kr{\kappa_{\rho}}
\def\aphi{\alpha_{\phi}}
\def\kphi{\kappa_{\phi}}
\def\3P0{$^3$P$_0$}

\title{\Large
Meson Emission Model of $\Psi \to N\bar Nm$ Charmonium Strong Decays}

\author{
T.Barnes,$^{a,b}$\footnote{Email: tbarnes@utk.edu}
Xiaoguang Li$^{b}$\footnote{Email: xli22@utk.edu}
and
W.Roberts$^{c}$\footnote{Email: wroberts@fsu.edu}}

\affiliation{
$^a$Physics Division, Oak Ridge National Laboratory,
Oak Ridge, TN 37831-6373, USA\\
$^b$Department of Physics and Astronomy, University of Tennessee,
Knoxville, TN 37996-1200, USA\\
$^c$Department of Physics and Astronomy, Florida State University,
Tallahassee, FL 32306-4350, USA}

\date{\today}

\begin{abstract}
In this paper we consider a sequential "meson emission" mechanism for charmonium 
decays of the type $\Psi \to N\bar Nm$, where $\Psi$ is a generic charmonium state,
$N$ is a nucleon and $m$ is a light meson.
This decay mechanism, which may not be dominant in general, assumes that an $N\bar N$ pair 
is created during charmonium annihilation, and the light meson $m$ is emitted 
from the outgoing nucleon or antinucleon line. A straightforward generalization 
of this model can incorporate intermediate $N^*$ resonances. We derive Dalitz plot event 
densities for the cases $\Psi = \eta_c$, $J/\psi$, $\chi_{c0}$, $\chi_{c1}$ and $\psi'$ 
and $m = \pi^0, f_0$ and $\omega$ (and implicitly, any $0^{-+}$, $0^{++}$ or $1^{--}$ 
final light meson). It may be possible to separate the contribution of this decay mechanism to the full 
decay amplitude through characteristic event densities. For the decay subset 
$\Psi \to p\bar p\pi^0$ the two model parameters are known, so we are able to 
predict absolute numerical partial widths for $\Gamma(\Psi\to p\bar p\pi^0)$.
In the specific case $J/\psi \to p\bar p\pi^0$ the predicted partial width and $M_{p\pi}$ event
distribution are intriguingly close to experiment. 
We also consider the possibility of scalar meson and glueball searches in $\Psi \to p\bar p f_0$.
If the meson emission contributions to $\Psi \to N\bar N m$ decays 
can be isolated and quantified, 
they can be used to estimate meson-nucleon strong couplings $\{g_{NNm}\}$, 
which are typically poorly known, and are a crucial input in meson exchange models of 
the $NN$ interaction. The determination of $g_{NN\pi}$ from $J/\psi \to p\bar p\pi^0$ 
and the (poorly known) $g_{NN\omega}$ and the anomalous "strong magnetic" coupling 
$\kappa_{NN\omega}$ from $J/\psi \to p\bar p\omega$ are considered as examples. 
\end{abstract}


\maketitle
\section{Introduction}
Charmonium strong decays of the type $\Psi \to \NNm$, where 
$\Psi$ is a generic charmonium state,
$N$ is a nucleon and $m$ is a light meson, 
have recently attracted interest both as sources of information regarding the $N^*$ spectrum 
\cite{Ablikim:2009iw,Ablikim:2009ay,Ablikim:2007dy,Ablikim:2005ir,Ablikim:2006aha}
and in searches for a low energy $\NN$ enhancement ``$X(1835)$", which has been reported in 
$J/\psi \to \gamma \pp$~\cite{Bai:2003sw}
and
$J/\psi \to \gamma \pi^+\pi^-\eta'$~\cite{Ablikim:2005um,YH_BES_HADRON2009}, 
but thus far not in $\Psi \to \ppm$.
These decays are also of interest because their partial widths can be used to estimate the 
$\pp \to m \Psi$ associated charmonium production cross sections at PANDA 
\cite{Lundborg:2005am,PandaTechnicalProgress}. As we shall show here, they may also provide information on
$NNm$ meson-nucleon coupling constants, which could be used to identify unusual resonances 
such as molecule or glueball candidates.

Specific $\Psi \to \NNm$ reactions that have recently been studied experimentally include 
$J/\psi \to \pppi$~\cite{Ablikim:2009iw},
$\pp\eta$ and $\pp\eta'$~\cite{Ablikim:2009ay}, 
and
$\ppw$~\cite{Ablikim:2007dy};
$\psi' \to \pppi$~\cite{Ablikim:2005ir},
$\ppeta$~\cite{Ablikim:2005ir,Briere:2005rc},
$p\bar n \pi^- + h.c.$~\cite{Ablikim:2006aha},
$\pp\rho$~\cite{Briere:2005rc} 
and $\ppw$~\cite{Briere:2005rc,Bai:2002yn}, 
and
(upper limit) $\pp\phi$~\cite{Briere:2005rc,Bai:2002yn};
and $\chi_{cJ} \to \pppi$ and $\pp\eta$~\cite{Athar:2006gq}.

\begin{figure}[b]
\vskip -4mm
\includegraphics[width=0.8\linewidth]{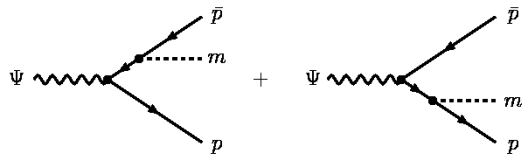}
\vskip -7cm
\caption{Feynman diagrams of the meson emission model.}
\label{fig:blr_fig1}
\end{figure}

These decays may prove to be complicated processes in which several decay mechanisms contribute significantly. 
For this reason it will be useful to have predictions for $\Psi \to \NNm$ Dalitz plot (DP) event 
densities assuming various decay mechanisms; this paper provides these results for one such mechanism. 
In particular we derive the DP event densities that follow from sequential meson emission,
in which the charmonium state (generically $\Psi$) decays to an intermediate $\NN$ state,
which radiates the light meson from the $N$ or $\bar N$ line, $\Psi \to \NN \to \NNm$.
The two Feynman diagrams assumed in this model are shown in Fig.\ref{fig:blr_fig1}.

We emphasize that the actual relative importance of this and other $\Psi \to \NNm$ 
decay mechanisms is unclear at present, and may depend strongly on the 
charmonium state $\Psi$ and the light meson $m$; one purpose of this paper is to determine 
the rates predicted by this meson emission decay model in isolation for comparison with experiment, 
so that the importance of this decay mechanism can be estimated.

The predictions of this $\Psi \to \NN \to \NNm$ decay model can be given in some cases 
with no free parameters, since the strengths of the {\it a priori} unknown couplings 
$\Psi\NN$ and $NNm$ can be estimated from other processes. Here we will 
give absolute predictions for the set of partial widths $\{\Gamma(\Psi \to \pppi)\}$; we
use the known partial widths $\{ \Gamma(\Psi \to \pp)\}$ to estimate the $\{ \Psi\NN\}$ couplings, 
and the final $NN\pi$ coupling is of course well known.

Provided that the contribution of the $\Psi \to \NN \to \NNm$ decay mechanism can be
isolated and quantified experimentally, this information can be used to estimate meson-nucleon 
strong coupling constants; these are generally poorly known, and play an important role 
in nuclear physics as input parameters in meson exchange models of the $NN$ force 
\cite{Machleidt:1987hj,Machleidt:2000ge,Nagels:1976xq,Nagels:1978sc,Stoks:1994wp,Cottingham:1973wt,
Lacombe:1980dr,Downum_thesis}. 

We will also discuss the 
determination of the (well known) $\pppi$ and the (poorly known) $\ppw$ couplings from the
decays $J/\psi \to \pppi$, $\psi' \to \pppi$ and $J/\psi \to \ppw$ as examples. 
This provides a third motivation for the study of $\Psi \to \NNm $ decays; 
they may prove useful for estimating $NNm$ coupling constants, 
in addition to their relevance to $N^*$ spectroscopy  
\cite{Ablikim:2009iw,Ablikim:2009ay,Ablikim:2007dy,Ablikim:2005ir,Ablikim:2006aha}
and low-mass $p\bar p$ dynamics \cite{Bai:2003sw,Ablikim:2005um,YH_BES_HADRON2009}. 
Another motivation for studying $\Psi \to \NNm$ is the possibility of observing light scalars, 
including the ``$\sigma$", the 980~MeV states and the scalar glueball, in the decays
$\Psi \to \ppf0$ (and $a_0$).

\section{Formulas}

Here we will usually specialize to charmonium decays to a 
$\pp$ pair and a neutral meson, $\Psi \to \ppm^0$; these decays are reasonably well studied, and
enjoy the simplification of equal baryon and antibaryon masses.
Our results employ conventions for kinematic variables, meson-baryon couplings and masses 
that were used in Ref.\cite{Barnes:2006ck}. In particular, $M$ is the mass of the initial 
charmonium state, $m_p$ is the proton mass, $m_m$ is the mass 
of meson (subscript) $m$, and dimensionless mass ratios $R \equiv M/\m$ and $r \equiv m_m/\m$ 
are defined relative to the proton mass. 
(Hence the numerical values of $R$ and $r$ depend on the decay process.)
We use scaled dimensionless variables $x = M_{pm}^2/\m^2 - 1$ and $y = M_{{\bar p}m}^2/\m^2 - 1$ 
and their inverses $u = 1/x$ and $v = 1/y$ to describe Dalitz plots; these greatly 
simplify our results. The DP event densities we present here are formally partial width densities 
in $x$ and $y$, which are related to the more familiar differential partial widths by a trivial overall constant,

\be
\frac{d^2\Gamma(\Psi \to \ppm)}{dxdy} 
=
m_p^4 \; \frac{d^2\Gamma(\Psi \to \ppm)}{dM^2_{pm} dM^2_{\bar pm}}\ .
\label{eq:DP_densities_defn}
\ee

Before we give our results for these event densities, it is useful 
to recall some general properties of a $\Psi \to \ppm$ 
Dalitz plot. The boundary in the dimensionless variables $(x,y)$ 
is specified by the curves
\be
y_{\pm} = 
\frac{
r^2 R^2  
+ 
(r^2 + R^2 - 2)x - x^2 
\pm
F_m F_{\Psi}}
{2(1+x)}
\label{eq:DP_boundaries}
\ee 
where $F_m = F(r,x)$ and $F_{\Psi} = F(R,x)$, with
\be
F(a,x) \equiv (a^2 (a^2 -4) - 2a^2 x + x^2)^{1/2}.
\label{eq:DP_boundaries_suppl}
\ee
The range of values of $x$ (and $y$) in the physical region is 
\be
r\, (r+2)\leq x \leq R\, (R-2).
\label{eq:DP_xrange}
\ee
The areas $\{ A_D\} $ of these Dalitz plots are useful for estimating 
$\Gamma(\Psi \to \ppm)$ partial widths \cite{Lundborg:2005am}. 
Although $A_D$ can be evaluated in closed form for $\Psi \to \ppm$ 
with general mass ratios $r$ and $R$, the resulting expression 
is quite lengthy, so when required we will simply evaluate 
each $A_D$ numerically.

In deriving the DP event densities we have usually assumed that the $\Psi\pp$ coupling is a 
constant $g_{\Psi\pp}$ times the simplest relevant Dirac matrix for the given 
$\Psi$ quantum numbers; for example, for the $J/\psi$ we use a pure vector $J/\psi\pp$ vertex, 
$-ig_{J/\psi\pp} \gamma_{\mu}$. The order of the hadron labels in $g_{abc}$ is meant to reflect 
the fact that the numerical value of this coupling constant is taken from an $a\to bc$ transition, 
here $J/\psi \to \pp$. This could be a significant concern if form factor effects are large.  

We proceed similarly for the light mesons $\pi^0$ and $f_0$;
for the pion we use a pure pseudoscalar $NN\pi$ coupling, with vertex $g_{NN\pi} \gamma_5$, 
and $-i g_{NNf_0} I$ for the $NNf_0$ vertex. Since light vector mesons (generically represented 
by the $\omega$) have two interesting 
strong couplings, Dirac (vector) and Pauli (anomalous magnetic), 
for this special case we assume a vertex with two interactions,
\be
\Gamma^{(\omega )}_{\mu} = 
-ig_{NN\omega}\bigg( \gamma_{\mu} + i (\kappa_{NN\omega}/2\m)\sigma_{\mu\nu} q_{\nu}\bigg)\,  .
\label{eq:omega_vertex}
\ee
Ref.\cite{Barnes:2007ub} assumed a similar $J/\psi\pp$ vertex; see Ref.\cite{Barnes:2006ck} 
for additional details regarding the couplings assumed here. 
We generally abbreviate these coupling constants as 
$g_{\Psi}\equiv g_{\Psi\NN}$ and $g_m \equiv g_{NNm}$; rationalized 
squared couplings $\alpha_{\Psi} \equiv g_{\Psi}^2/4\pi$ and 
$\alpha_m \equiv g_m^2/4\pi$ are also used. 

For the special case of $\pppi$ final states, these event densities
can be obtained by applying crossing relations to our previously 
published results for the unpolarized differential cross sections 
for the $2\to 2$ processes $\pp \to \pi^0 \Psi$ \cite{Barnes:2006ck}; 
the other ($f_0$ and $\omega$) cases have not been considered previously. 
The results for all cases considered here are given below. 
\ba
&&
\frac
{d^2\Gamma({\eta_c \to \pppi})}
{dxdy} =  
\alpha_{\eta_c}\, \alpha_{\pi}
\;
\frac
{\m}{8 \pi R^3}\; 
\Bigg\{
(u-v)^2
\cdot
\nonumber\\
&&
\Big(
\frac{1}{uv} - r^2 R^2
\Big)
\Bigg\} 
\label{eq:etac_pppi0}
\\
\nonumber\\
&& 
\frac
{d^2\Gamma({J/\psi \to \pppi})} 
{dxdy} =  
\alpha_{J/\psi}\, \alpha_{\pi}
\;
\frac
{\m}{12 \pi R^3}\; 
\Bigg\{
\frac{(u+v)^2}{uv}
\nonumber\\
&&
- 2(u+v) (u+v+1) \, r^2 
+ 2uv \, r^4
- (u^2+v^2) \cdot 
\nonumber\\
&&
r^2 R^2 
\Bigg\} 
\label{eq:Jpsi_pppi0}
\\
\nonumber\\
&&
\frac
{d^2\Gamma({\chi_{c0} \to \pppi})} 
{dxdy} 
=  
\alpha_{\chi_{c0}}\, \alpha_{\pi}
\;
\frac{\m}{8 \pi R^3}\; 
\Bigg\{
(u+v)^2
\cdot
\nonumber\\
&&
\Big(
\frac{1}{uv} + 4\, r^2 - r^2 R^2
\Big)
\Bigg\}
\label{eq:chic0_pppi0}
\\
\nonumber\\
&&
\frac
{d^2\Gamma({\chi_{c1} \to \pppi})} 
{dxdy} =  
\alpha_{\chi_{c1}}\, \alpha_{\pi}
\;
\frac{\m}{6 \pi R^5}\, 
\Bigg\{
-\frac{(u+v)}{uv} \cdot
\nonumber\\
&&
(u+v+1) 
+  r^2
+ \frac{(u^2+v^2)}{2uv} \, R^2 
+\Big( 2(u^2+v^2) 
\nonumber\\
&&
+ u + v \Big) \, r^2 R^2 
-uv \, r^4 R^2 
- \frac{(u^2+v^2)}{2}  \, r^2 R^4 
\Bigg\}
\label{eq:chic1_pppi0}
\\
\nonumber\\
&&
\frac
{d^2\Gamma({\eta_c \to \ppf0})} 
{dxdy} =
\alpha_{\eta_c}\, \alpha_{f_0}
\;
\frac
{\m}{8 \pi R^3}\; 
\Bigg\{
(u+v)^2 \cdot
\nonumber\\
&&
\Big(
\frac{1}{uv} 
+ 4 R^2 - r^2 R^2
\Big)
\Bigg\} 
\label{eq:etac_ppf0}
\\
\nonumber\\
&& 
\frac
{d^2\Gamma({J/\psi \to \ppf0})} 
{dxdy} = 
\alpha_{J/\psi}\, \alpha_{f_0}
\;
\frac
{\m}{12 \pi R^3}\; 
\Bigg\{
\frac{(u+v)^2}{uv} 
\nonumber\\
&&
+ 8 (u+v)(u+v+1) -2\Big( u(u+1) + v(v+1) 
\nonumber\\
&&
+ 6uv \Big) \, r^2
+ 4(u+v)^2\, R^2 
+ 2uv \, r^4 
- (u^2+v^2) \cdot
\nonumber\\
&&
r^2 R^2
\Bigg\} 
\label{eq:Jpsi_ppf0}
\\
\nonumber\\
&&
\frac
{d^2\Gamma({\chi_{c0} \to \ppf0})} 
{dxdy} = 
\alpha_{\chi_{c0}}\, \alpha_{f_0}
\;
\frac
{\m}{8 \pi R^3}\; 
\Bigg\{
\frac{(u-v)^2}{uv}
\nonumber\\
&&
- 16(u+v)(u+v+1) + 4 (u+v)^2 (r^2 + R^2) 
\nonumber\\
&&
- (u-v)^2 \, r^2 R^2
\Bigg\} 
\label{eq:chic0_ppf0}
\\
\nonumber\\
&&
\frac
{d^2\Gamma({\chi_{c1} \to \pp f_0})} 
{dxdy} =  
\alpha_{\chi_{c1}}\, \alpha_{f_0}
\;
\frac{\m}{6 \pi R^5}\, 
\Bigg\{
-\frac{(u+v)}{uv} \cdot
\nonumber\\
&&
(u+v+1) 
+  r^2
+ 
\Big( \frac{(u^2+v^2)}{2uv} - 8(u+v)\cdot 
\nonumber\\
&&
(u+v+1) \Big) \, 
R^2 
+\Big( 2(u^2+v^2) + 8uv + u + v \Big) \, r^2 R^2 
\nonumber\\
&&
+ 2 (u+v)^2 \, R^4  
-uv \, r^4 R^2 
- \frac{(u^2+v^2)}{2}  \, r^2 R^4 
\Bigg\}
\label{eq:chic1_ppf0}
\\
\nonumber\\
&&
\frac
{d^2\Gamma({\eta_c \to \ppw})} 
{dxdy} = 
\alpha_{\eta_c}\, \alpha_{\omega}
\;
\frac
{\m}{4 \pi R^3}\; 
\Bigg\{
\bigg[
\frac{(u+v)^2}{uv} 
\nonumber\\
&&
- 2(u+v)(u+v+1)\, R^2 
- (u^2+v^2)\, r^2R^2
+ 2uv\, R^4 
\bigg]
\nonumber\\
&&
+ \kappa_{\omega} 
\bigg[ 
- \frac{2(u+v)^2}{uv} 
+ \Big( 3(u^2+v^2) + 2 uv \Big)\, r^2 R^2 
\bigg]
\nonumber\\
&&
+ \kappa_{\omega}^2 
\bigg[
\frac{(u+v)(u+v-1)}{2uv} 
+ \frac{(u+v)^2}{8uv} \, r^2  
+ \frac{1}{2}\, R^2 
\nonumber\\
&&
+ \bigg( \frac{(u+v)}{2} -( u^2 + v^2) \bigg) \, r^2 R^2
- \frac{(u+v)^2}{8} r^4 R^2 
\nonumber\\
&&
-\frac{uv}{2} r^2 R^4 
\bigg]
\Bigg\}
\label{eq:etac_ppw}
\\
\nonumber\\
&&
\frac
{d^2\Gamma({J/\psi \to \ppw})} 
{dxdy} = 
\alpha_{J/\psi}\, \alpha_{\omega}
\;
\frac
{\m}{6 \pi R^3}\; 
\Bigg\{
\bigg[
\frac{(u^2+v^2)}{uv} 
\nonumber\\
&&
- 4(u+v)(u+v+1)
- 2 \Big(u(u+1) +v(v+1)\Big)\cdot
\nonumber\\
&&
(r^2 + R^2) 
+ 2 uv\, r^4 
- (u^2 + v^2 - 4uv) \, r^2 R^2
+ 2 uv\, R^4 
\bigg]
\nonumber\\
&&
+ \kappa_{\omega} \bigg[ - \frac{(u+v)^2}{uv} + 6 (u+v)(u+v+1) \, r^2  
- 6 uv\, r^4 
\nonumber\\
&&
+ \Big( 3(u^2+v^2) - 8 uv \Big)\, r^2 R^2  
\bigg]
+ \kappa_{\omega}^2 \bigg[
- \frac{(u+v)}{2uv} 
\nonumber\\
&&
+ \Big( 
\frac{(u+v)^2}{8uv} 
- 2(u+v)(u+v+1) \Big) \, r^2  
+ \frac{1}{2}\, R^2 
\nonumber\\
&&
- \frac{1}{4} \Big( u(u+1) + v(v+1) 
- 6 uv \Big) \, r^4  
- (u-v)^2\, r^2R^2 
\nonumber\\
&&
+ \frac{1}{4} uv\, r^6 
-\frac{1}{8} (u^2+v^2-4uv)\, r^4 R^2 
\bigg]
\Bigg\}
\label{eq:Jpsi_ppw}
\\
\nonumber\\
&&
\frac
{d^2\Gamma(\chi_{c0} \to \ppw)} 
{dxdy} = 
\alpha_{\chi_{c0}}\, \alpha_{\omega}
\;
\frac{\m}{4 \pi R^3}\; 
\Bigg\{
\bigg[
\frac{(u+v)^2}{uv} 
\nonumber\\
&&
+ 8(u+v)(u+v+1)
+ 4 (u+v)^2\, r^2 
- 2 \Big( u(u+1)
\nonumber\\
&&
+v(v+1) + 6uv \Big)\, R^2
- (u^2+v^2) \, r^2 R^2
+ 2 uv\, R^4 
\bigg]
\nonumber\\
&&
+ \kappa_{\omega} \bigg[ - 12 (u+v)(u+v+\frac{1}{2}) \, r^2  
+  3(u+v)^2 \, r^2 R^2  \bigg]
\nonumber\\
&&
+ \kappa_{\omega}^2 \bigg[
- \frac{(u+v)(u+v+1)}{2uv} 
+ \bigg( \frac{(u^2+6uv+v^2)}{8uv} 
\nonumber\\
&&
+ 4(u+v)(u+v+1) \bigg)\, r^2
+ \frac{1}{2}\, R^2 
+ \frac{(u+v)^2}{2}\, r^4 
\nonumber\\
&&
- \bigg( u^2 + 4uv + v^2 + \frac{(u+v)}{2}\bigg) \, r^2R^2 
- \frac{(u-v)^2}{8}\,r^4 R^2
\nonumber\\
&&
+ \frac{uv}{2}\, r^2R^4 
\bigg]
\Bigg\}
\label{eq:chic0_ppw}
\\
\nonumber\\
&&
\frac
{d^2\Gamma({\chi_{c1} \to \ppw})} 
{dxdy} =  
\alpha_{\chi_{c1}}\, \alpha_{\omega}
\;
\frac{\m}{3 \pi R^5}\, 
\Bigg\{
\bigg[
\frac{(u+v)^2}{uv} 
\nonumber\\
&&
+ \bigg(
\frac{(u^2+v^2)}{2uv} + 4(u+v)(u+v+1) 
\bigg) \, R^2
\nonumber\\
&&
+ \Big( 2(u-v)^2 
- u - v \Big) \, r^2R^2 
- ( u^2 + v^2 
+ 6uv 
\nonumber\\
&&
+ u + v)\, R^4
+ uv\, r^4 R^2  
+\Big( 2uv - \frac{(u^2+v^2)}{2} \Big)\, r^2R^4 
\nonumber\\
&&
+ uv\, R^6 
\bigg]
+ \kappa_{\omega} 
\bigg[
- \frac{(u+v)^2}{uv} - r^2 
- \frac{(u^2+v^2-4uv)}{2uv}\cdot 
\nonumber\\
&&
R^2
- \Big( 6(u^2+v^2) + u + v\Big) \, r^2 R^2
- uv\, r^4 R^2 
\nonumber\\
&&
+ \frac{3}{2} (u^2+v^2) \, r^2 R^4 
\bigg]
+ \kappa_{\omega}^2 
\bigg[
\frac{(1 +2(u+v) + 2(u+v)^2)}{4uv}
\nonumber\\
&&
- \frac{( u + v + (u-v)^2)}{8uv}\, r^2
- \bigg( \frac{3}{2}+\frac{(u+v)}{4uv} \bigg) \, R^2
+ \frac{1}{8}\, r^4   
\nonumber\\
&&
+ \bigg( 2(u^2 + v^2 - uv) 
+ \frac{(u+v)}{2} 
+ \frac{(u^2 + v^2 + 4uv)}{16uv}
\bigg)\cdot  
\nonumber\\
&&
r^2 R^2
+ \frac{1}{4}\,R^4  
+\frac{1}{8}\Big( u + v + 2(u^2 + v^2 + 6uv) \Big)\,  
r^4 R^2
\nonumber\\
&&
-\frac{1}{2} (u^2 + v^2 - uv) \,r^2R^4
- \frac{1}{8}uv\, r^6 R^2
\nonumber\\
&&
- \frac{1}{16} (u^2 +v^2 + 4uv) \,r^4 R^4
\bigg]
\Bigg\}\; .
\label{eq:chic1_ppw}
\ea

The symmetry of these event densities under $(x,y)$ and hence $(u,v)$ 
interchange is a consequence of $C$-parity invariance. 
There are singularities in these events densities along the lines 
$x=0$ ($u = \infty$) and $y=0$ ($v = \infty$), corresponding to $M_{pm}^2 = m_p^2$ and 
$M_{{\bar p}m}^2 = m_p^2$. These are due to the poles of 
the $p$ and $\bar p$ propagators in the Feynman diagrams 
for the decay process $\Psi \to \pp \to \ppm$ (Fig.\ref{fig:blr_fig1})
and lie outside the physical regions of the Dalitz plots. 

\section{Applications}

\subsection{The $J/\psi \to \pppi$ partial width}

As a first application we will evaluate the partial width for $J/\psi \to \pppi$ assuming 
this meson emission decay mechanism. The PDG \cite{PDG} 
quotes a branching fraction of
\be
B(J/\psi \to \pppi) = 1.09 \pm 0.09 \cdot 10^{-3}
\label{eq:Jpsi_pppi_bf_pdg} 
\ee
which is the average of early measurements by 
Mark-I \cite{Peruzzi:1977pb}, DASP \cite{Brandelik:1979hy} and Mark-II \cite{Eaton:1983kb}.
There are also more recent experimental results on this decay from 
BES-II \cite{Ablikim:2009iw}.
Using the current PDG value for the $J/\psi$ total width of $93.2\pm 2.1$~keV, 
this branching fraction corresponds to a partial width of
\be
\Gamma(J/\psi \to \pppi) = 102 \pm 9 \; \hbox{eV}.
\label{eq:Jpsi_pppi_width_pdg} 
\ee

\begin{figure}[t]
\vskip -6mm
\includegraphics[width=0.8\linewidth]{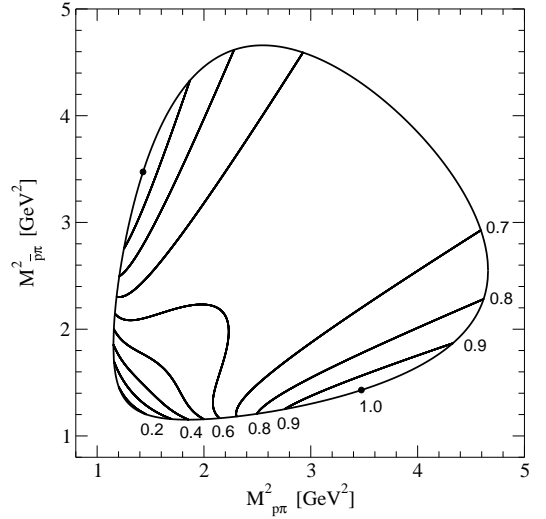}
\caption{The $J/\psi\to\pppi$ DP event density predicted 
by the meson emission decay mechanism $J/\psi\to\pp\to\pppi$, 
Eq.(\ref{eq:Jpsi_pppi0}). 
Contours of equal density are shown.}
\label{fig:blr_fig2}
\end{figure}

To evaluate this partial width in the meson emission model we simply integrate the theoretical 
event density (\ref{eq:Jpsi_pppi0}) over the Dalitz plot. This event density is given by
\be
\frac{d^2\Gamma({J/\psi \to \pppi})}{dxdy} = 
\alpha_{J/\psi}\, \alpha_{\pi}
\, \m \, \rho (x,y) 
\label{eq:Jpsi_pppirate}
\ee
where the dimensionless density function $\rho$ is
\bd
\rho (x,y) = 
\frac
{1}{12 \pi R^3}\; 
\bigg\{
\frac{(u+v)^2}{uv}
-2(u+v)(u+v+1) \, r^2
\ed
\be
+ 2 uv \, r^4 
- (u^2+v^2) \, r^2  R^2
\bigg\}\; . 
\label{eq:rho_Jpsi_pppirate}
\ee
This (parameter-free) relative event density, scaled to the maximum value
in the physical region, is shown in Fig.\ref{fig:blr_fig2}.
Integrating (\ref{eq:Jpsi_pppirate}) over the Dalitz plot gives
\be
\Gamma({J/\psi \to \pppi}) = \alpha_{J/\psi}\, \alpha_{\pi}\, m_p  \cdot  
\langle \rho \rangle \cdot A_D / m_p^4\, . 
\label{eq:3body_rate}
\ee 
where $\langle \rho \rangle$ is the mean value of $\rho(x,y)$ in the Dalitz plot, 
which has (physical) area $A_D$; 
\be
\int\!\!\!\!\int_{DP}\;  \rho\; dx\, dy = \langle \rho \rangle \cdot A_D / m_p^4 \, . 
\label{eq:DP_int_defn}
\ee  
We evaluate $\langle \rho \rangle$ and $A_D$ numerically, 
assuming physical hadron masses; we use PDG masses rounded to 0.1~MeV;  
$m_{\pi^0} = 0.1350$~GeV, $\m   = 0.9383$~GeV and $m_{J/\psi}= 3.0969$~GeV, 
which leads to
$\langle \rho \rangle = 3.070 \cdot 10^{-3}$ and $A_D = 9.265$~GeV$^4$,
and a partial width of
\be
\Gamma({J/\psi \to \pppi}) = 34.44 \cdot \alpha_{J/\psi}\, \alpha_{\pi}\; \hbox{MeV}.
\label{eq:3body_rate_num1}
\ee 
To complete this estimate we require numerical values for the $NN\pi$ and 
$J/\psi \NN$ coupling constants. For $NN\pi$ there is general agreement 
from meson exchange models of $NN$ scattering that $g_{NN\pi} \approx 13$
(see for example 
\cite{Machleidt:1987hj,Machleidt:2000ge,Nagels:1976xq,Nagels:1978sc,Stoks:1994wp,Cottingham:1973wt,
Lacombe:1980dr,Downum_thesis}); 
we accordingly set $g_{NN\pi} = 13.0$. 
The value of the $J/\psi\NN$ coupling constant (here $g_{J/\psi\pp}$) can be 
estimated from the measured partial width to $p\bar p$, which is (again using PDG numbers)
$\Gamma(J/\psi \to p\bar p) = 202 \pm 8 $~eV. Equating this to the theoretical decay rate 
\be
\Gamma({J/\psi \to p\bar p}) 
=  \ \alpha_{J/\psi}\, \beta_p
( 1 + 2/R^2 ) M / 3
\label{eq:Jpsi_ppbarwidth_formula}
\ee
gives a value of 
$g_{J/\psi\pp} = 1.62\cdot 10^{-3}$, as was quoted previously in Ref.\cite{Barnes:2006ck}. 
Using these couplings in Eq.(\ref{eq:3body_rate_num1}) gives our meson emission model prediction 
\be
\Gamma({J/\psi \to \pppi}) = 97 \; \hbox{eV}.
\label{eq:3body_rate_num2}
\ee 
This is consistent with the experimental value of $102\pm 9$~eV quoted in 
Eq.(\ref{eq:Jpsi_pppi_width_pdg}). This excellent agreement is somewhat fortuitous, 
since this version of the model does not include the $N^*$ contributions evident in the
$J/\psi \to \pppi$ Dalitz plot \cite{Ablikim:2009iw} (see also Fig.\ref{fig:blr_fig3}). 

We note in passing that the charged-pion cases $J/\psi \to p \bar n \pi^-$ and 
$n\bar p \pi^+$ should have branching fractions 
close to twice $B(J/\psi \to \pppi)$, reflecting an isospin factor of two. 
Experimentally this is indeed the case; the ratio of 
the PDG $J/\psi$ branching fractions to $p \bar n \pi^-$ and $\pppi$ is 
$B(J/\psi \to p \bar n \pi^-)/B(J/\psi \to \pppi) = 1.94 \pm 0.18$. 

\subsection{Projected $J/\psi \to \pppi$ event densities}

Projections of DP event densities onto the invariant mass of one pair of particles 
are useful in searches for intermediate resonances. For $J/\psi \to \pppi$, the BES 
Collaboration has published acceptance-corrected event densities in $M_{p\pi}$ and $M_{\bar p\pi}$
invariant mass (Fig.6 of Ref.\cite{Ablikim:2009iw}), which show clear evidence for $N^*$ resonances.
Here we will generate the corresponding theoretical $M_{p\pi}$-projected event distributions 
in the meson emission model for comparison with experiment. Although $N^*$ resonances are 
not incorporated in our calculation, this comparison will test the relative importance of 
the meson emission decay mechanism in this decay, and establish whether the model predicts 
a non-$N^*$ ``background" invariant mass distribution that is similar to the data in form 
and magnitude. 

\begin{figure}[t]
\includegraphics[width=0.9\linewidth]{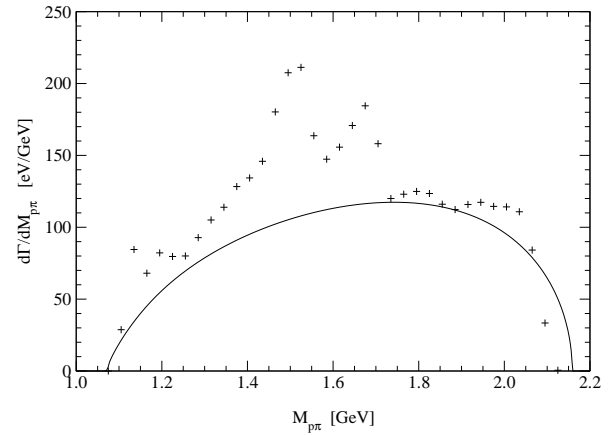}
\caption{$J/\psi\to\pppi$ experimental (BES) 
and theoretical (meson emission model, Fig.\ref{fig:blr_fig2}) Dalitz plot (DP) event densities, 
projected onto $M_{p\pi}$. This is not a fit; see text for discussion.}
\label{fig:blr_fig3}
\end{figure}

The full two-dimensional DP event density $d^2\Gamma/dxdy$ predicted by the meson emission 
model is given by Eq.(\ref{eq:Jpsi_pppi0}). Projecting this onto $M_{p\pi}$ 
is straightforward; first one integrates over $y = M_{\bar p\pi}^2/\m^2 - 1$ between 
the DP boundaries $y_{\pm}(x)$ of Eq.(\ref{eq:DP_boundaries}), which gives $d\Gamma/dx$.
Converting this into a distribution in $M_{p\pi}$ introduces a Jacobean, which is specified by the definition
$x = M_{p\pi}^2/\m^2 - 1$. This gives
\be
\frac{d\Gamma(J/\psi \to \pppi)}{dM_{p\pi}} = 
\frac{2M_{p\pi}}{\m^2} 
\int^{y_+(x)}_{y_-(x)} dy\;
\frac{d^2\Gamma(J/\psi \to \pppi)}{dxdy}\; .
\label{eq:Projected_mass_distn}
\ee
We have evaluated this distribution numerically, given the $d^2\Gamma(J/\psi \to \pppi)/dxdy$ 
of Eq.(\ref{eq:Jpsi_pppi0}) and the masses and couplings $\alpha_{J/\psi}$ and $\alpha_{\pi}$
used in Sec.III.A. The result is shown in Fig.\ref{fig:blr_fig3}, together with an experimental 
distribution provided by BES (reported in Ref.\cite{Ablikim:2009iw}), using a common scale. 
The data is the combined acceptance-corrected $M_{p\pi}$ and $M_{\bar p\pi}$ distribution, 
scaled to give their reported $B(J/\psi \to \pppi) = 1.33\cdot 10^{-3}$ 
rather than the PDG value of $1.09\cdot 10^{-3}$.

Clearly there is a close resemblance between the meson emission model prediction for the 
$J/\psi \to \pppi$ event distribution in $M_{p\pi}$ and the observed BES distribution, both in 
form and magnitude. This suggests that a study of this reaction assuming this model for the 
experimental ``background" combined with $N^*$ resonance contributions would be an interesting 
exercise. Although BES~\cite{Ablikim:2009iw} recently reported a similar study, they introduced
an {\it ad hoc} $s_{\pi N} (= M_{p\pi}^2)$-dependent form factor that suppressed this ``background" 
meson emission amplitude relative to $N^*$ contributions. The similarity to experiment 
evident in~Fig.\ref{fig:blr_fig3} suggests that this mechanism merits additional consideration. 

\subsection{Other $\Psi \to \pppi$ partial widths}

Since the two meson emission model parameters $g_{\Psi\pp}$ and $g_{NNm}$ are both known for several 
$\Psi \to \pppi$ decays, we are able to give absolute predictions for these partial widths. 
(We previously used $\Gamma(\Psi \to \pp)$ to estimate $g_{\Psi\pp}$ \cite{Barnes:2006ck};
here we again use these values, and set $g_{NN\pi}=13$.) 
These $\Psi \to \pppi$ partial widths are given in Table~\ref{table:pppi_rates}, together with some 
intermediate theoretical quantities and the experimental widths. (These experimental values are the
PDG total widths times branching fractions, with errors added in quadrature.)

\begin{table}[h]
\vskip 3mm
\begin{center}
\begin{tabular}{|c|c|c|c|r|r|}
\hline
$\Psi$
& $10^3 \cdot g_{\Psi \pp}$
& $10^3 \cdot \langle\rho\rangle $
& $A_D$~[GeV$^4$]
& $\Gamma^{thy.}_{\pppi}$\phantom{x,}
& $\Gamma^{expt.}_{\pppi}$\phantom{xxx}
\\
\hline
\
$\eta_c$
& 19.0\phantom{00}
& 0.530
& \phantom{0}6.862
& 1.7 keV
& -\phantom{xxxxxx}
\\
$J/\psi$
& 1.62
& 3.070
& \phantom{0}9.265
& 97 eV 
& $102\pm 9$ eV 
\\
$\chi_0$
& 5.42
& 3.691
& 18.605
& 2.6 keV
& $6.0\pm 1.3$ keV
\\
$\chi_1$
& 1.03
& 0.554
& 22.351
& 17 eV
& $103\pm 43$ eV
\\
$\psi'$
& 0.97
& 2.010
& 30.501
& 75 eV
& $41\pm 5$ eV
\\
\hline
\end{tabular}
\caption{Comparison of theory (meson emission model) and experiment for 
$\Gamma(\Psi \to \pppi)$ (see text).}
\label{table:pppi_rates}
\end{center}
\vskip -2mm
\end{table}
                                                                        
These rates were derived using Eq.(\ref{eq:3body_rate}), with the appropriate density function 
$\rho$ chosen from the set~Eqs.(\ref{eq:etac_pppi0}-\ref{eq:chic1_pppi0}). 
In addition to the rates, Table~\ref{table:pppi_rates} 
also gives the coupling constants assumed, the average of the density function $\rho$ over the Dalitz plot,
and the DP area $A_D$ in physical units. 

It is clear from the table that the wide variation in the absolute scale of 
partial widths observed experimentally is approximately reproduced by the model, 
at least at a ``factor-of-two" level of accuracy. This suggests that the meson emission decay 
mechanism may indeed be an important component of the decay amplitude in all these decays; a more definitive 
test would involve a direct comparison of the DP event densities or their two-body projections,
as in Fig.\ref{fig:blr_fig3}.

The $\chi_{c1}$ case appears to be an exception to this approximate agreement, 
however in view of the large experimental error it is not clear if this is a real discrepancy; 
theory and experiment only differ by $2\sigma$.

Although the single experimentally unobserved decay $\eta_c\to\pppi$ is predicted
by the meson emission model to have a relatively large partial width of 1.7~keV, it is actually considerably
suppressed by the presence of an on-diagonal node in the DP event distribution. An experimental study of
$\eta_c\to\pppi$ would accordingly be very interesting, since the contributions of other decay mechanisms 
may be easier to identify in the region of the DP where the meson emission model gives a zero or weak contribution.  

\subsection{$g_{NN\pi}$ from $B(J/\psi \to \pppi)/B(J/\psi \to \pp)$}

Previously we noted that $\Psi \to \ppm$ decays can be used to estimate 
$NNm$ couplings, provided that the contribution of the meson emission decay 
mechanism to the decay amplitude can be quantified experimentally. In the following 
we will use the decay $J/\psi \to \pppi$ as an illustration of this approach, since 
the agreement between the experimental and theoretical partial widths suggests that 
domination of this decay by meson emission is a reasonable first approximation.

Since the {\it a priori} unknown coupling $\alpha_{J/\psi}$ cancels in the 
theoretical branching fraction ratio $B(J/\psi \to \pppi)/B(J/\psi \to \pp)$, we can use it 
to estimate $g_{NN\pi}$ directly. The meson emission model decay width 
for $\Gamma(J/\psi \to \pppi)$ (\ref{eq:Jpsi_pppi0})
and the two-body decay width (\ref{eq:Jpsi_ppbarwidth_formula})
imply the following relation between this ratio and
the coupling $\alpha_{\pi} \equiv g_{NN\pi}^2/4\pi$:
\be
\alpha_{\pi} = 
(1-4/R^2)^{1/2}\; 
\frac{(R+2/R)}{3\langle \rho \rangle A_D / m_p^4}
\cdot
\frac{B(J/\psi \to \pppi)}{B(J/\psi \to \pp)}\; .
\label{eq:g_NNpi_fm_Jpsi_bfs}
\ee
Substitution of the experimental PDG numbers  
$B(J/\psi \to \pppi) = (1.09 \pm 0.09) \cdot 10^{-3}$ and 
$B(J/\psi \to \pp)   = (2.17 \pm 0.07) \cdot 10^{-3}$ 
for these branching fractions leads to the estimate
\be
g_{NN\pi}\bigg|_{J/\psi\to\pppi}  = 13.3 \pm 0.6
\label{eq:g_NNpi_fm_Jpsi_number}
\ee
which is consistent with $NN$ meson exchange model values.

We expect to find approximately equal $g_{NN\pi}$ estimates from other 
$\Psi \to \pp$ and $\ppm$ decay pairs if the meson emission decay mechanism 
$\Psi \to \pp \to \pppi$ is indeed dominant. A second state $\Psi$ that can be use to estimate 
$g_{NN\pi}$ is the $\psi'(3686)$. Since the $\psi'$ has the same quantum numbers as the 
$J/\psi$, Eq.(\ref{eq:g_NNpi_fm_Jpsi_bfs}) is again appropriate for our coupling constant 
estimate. This $\psi'$ case has a much larger $\pppi$ DP area $A_D$ than the 
$J/\psi$, which is partially compensated by a smaller mean event density $\langle \rho \rangle$. 
(These quantities are given in Table~\ref{table:pppi_rates}.) Using $M = 3.6861$~GeV and 
the PDG branching fractions 
$B(\psi' \to \pppi) = (1.33 \pm 0.17) \cdot 10^{-4}$ and 
$B(\psi' \to \pp)   = (2.75 \pm 0.12) \cdot 10^{-4}$, we find the $\psi'$-based $g_{NN\pi}$ estimate
\be
g_{NN\pi}\bigg|_{\psi'\to\pppi} = 9.9 \pm 0.7\ .
\label{eq:g_NNpi_fm_psip_number}
\ee
This is similar to but somewhat smaller than the estimate obtained above from $J/\psi$ decays
(\ref{eq:g_NNpi_fm_Jpsi_number}), and may give an indication of the accuracy of this 
approach for estimating $NNm$ coupling constants.

Of course the other relations for $\alpha_m$ analogous to (\ref{eq:g_NNpi_fm_Jpsi_bfs})
will only be useful if the meson emission decay mechanism is dominant in those decays as well. 
Otherwise the contribution of this mechanism to the decay must be identified and quantified, 
for example through a detailed study of the DP event density.

\subsection{Scalar mesons in $\Psi\to \NNm$}

The long-standing interest in the light scalars makes the possibility of studying them 
using these decays an attractive prospect. This motivated our inclusion of decay formulas 
for the processes $\Psi\to \pp f_0$ in our set of theoretical DP
event densities. 

Here we will give meson emission model predictions for the branching fractions 
$B(\Psi \to \pp f_0)$, where $\Psi = \eta_c, J/\psi, \chi_{c0}, \chi_{c1}$ and 
$\psi'$, for a light ``sigma" meson with $m_{f_0} = 0.5$~GeV. 
To evaluate these partial widths we proceed
as in Sec.III.C, and integrate the appropriate decay width formula from the set
Eqs.(\ref{eq:etac_ppf0}-\ref{eq:chic1_ppf0}) over the Dalitz plot. We again use the $\Psi\pp$ 
coupling constants of Sec.III.C, as given in Table~\ref{table:pppi_rates}. 
The total widths used to convert the calculated partial widths to branching fractions are the 
current PDG values,
$\Gamma(\eta_c) = 27.4$~MeV,
$\Gamma(J/\psi) = 93.2$~keV,
$\Gamma(\chi_{c0}) = 10.4$~MeV,
$\Gamma(\chi_{c1}) = 0.86$~MeV and
$\Gamma(\psi') = 309$~keV.
Our results are given in Table~\ref{table:ppf0_rates}.
Since there is no general agreement regarding an $NNf_0(500)$ coupling constant, in the table
we first give the predicted branching fraction relative to 
$B(J/\psi \to \pp f_0(500)) \equiv B_0$, which is numerically 
$0.338\cdot 10^{-4} \cdot g_{NNf_0}^2$.
(The unknown $g_{NNf_0}$ cancels in these ratios.) 
The next column gives absolute $B(\Psi \to \pp f_0(500))$ branching fractions for a rather arbitrarily chosen
$g_{NNf_0} = 10$. Finally, the table quotes experimental branching fractions for the related processes
$\Psi \to \pp \pi^+ \pi^-$ for comparison. 

The relative theoretical branching fractions 
(Table~\ref{table:ppf0_rates}, col.2) suggest that 
the best channel for identifying a light scalar in $\Psi \to \pp f_0$ is 
$J/\psi \to \pp f_0$ itself (assuming that the meson emission model is a reasonable guide).
Given a somewhat larger event sample, $\psi' \to \pp f_0$ should be competitive with
$J/\psi$, and has the advantage of more phase space, so the scalars near 1~GeV 
and the $f_0(1500)$ could also be investigated.
$\eta_c \to \pp f_0$ has a comparable theoretical branching fraction, but the difficulty of producing the 
$\eta_c$ makes this a less attractive channel. Finally, the $\chi_{cJ}$ states are predicted to
have much smaller $\ppf0$ branching fractions than $J/\psi \to \ppf0$, and accordingly are 
less attractive experimentally if this decay model is accurate.

\vskip 0.5cm
\begin{table}[h]
\begin{center}
\begin{tabular}{|c|c|c|c|}
\hline
$\Psi$
& $B^{thy.}_{\ppf0}/B_0$
& $10^3\cdot B^{thy.(g=10)}_{\pp f_0}$
& $10^3\cdot B^{expt.}_{\pp\pi^+\pi^-}$
\\
\hline
\
$\eta_c$
& $0.40$
& $1.4$
& $< 12$ (90\% c.l.)
\\
$J/\psi$
& $\equiv 1$ 
& $3.4$ 
& $6.0 \pm 0.5$ 
\\
$\chi_0$
& 0.045
& $0.15$
& $2.1\pm 0.7$
\\
$\chi_1$
& $0.016$
& $0.054$
& $0.50 \pm 0.19$
\\
$\psi'$
& $0.21$
& $0.72$
& $0.60 \pm 0.04$
\\
\hline
\end{tabular}
\caption{Theoretical (meson emission model) branching fractions 
for light scalar meson production. The numerical columns are
{\it 1)} The ratio $B(\Psi \to \pp f_0(500))/B(J/\psi \to \pp f_0(500))$; 
{\it 2)} $10^3\cdot B(\Psi \to \pp f_0(500))$ for $g_{NNf_0} = 10$; 
{\it 3)} $10^3\cdot B^{expt.}(\Psi \to \pp \pi^+\pi^-)$, for comparison with 
{\it 2)}.
(See text.)}
\label{table:ppf0_rates}
\end{center}
\end{table}
\vskip -0.5cm

We have also estimated the effect of an $f_0(500)$ width on these results. 
Imposing a Breit-Wigner $f_0$ mass profile with $\Gamma_{f_0} = 0.5$~GeV, truncated at $2m_{\pi}$, 
decreases all the absolute theoretical $\Psi \to p\bar p f_0(500)$ branching fractions 
in Table~\ref{table:ppf0_rates} (col.3) by $\approx 10\%$. The {\it relative} theoretical branching 
fractions (col.2) are even less sensitive to the $f_0(500)$ width, and become 
$0.41, 0.048, 0.017$ and $0.21$. 

A light scalar $f_0$ meson would presumably decay strongly and perhaps dominantly 
to $\pi\pi$, so decays of the type $\Psi \to \pp\pi\pi$ are of special interest, 
notably $J/\psi \to \pp\pi\pi$ (in view of our large theoretical 
$B(J/\psi \to \pp f_0(500))$). The charged case $J/\psi \to \pp\pi^+\pi^-$ has been 
studied by Mark-I~\cite{Peruzzi:1977pb}, DESY~\cite{Besch:1981ka}
and Mark-II~\cite{Eaton:1983kb}.
Although no light scalar mesons have yet been identified in this decay, 
it is suggestive that $J/\psi \to \pp\pi^+\pi^-$ is the largest known exclusive 
$J/\psi \to \pp X$ mode, with a branching fraction of 
$B(J/\psi \to \pp\pi^+\pi^-) = (6.0 \pm 0.5) \cdot 10^{-3}$. 

In addition to the $\pp f_0$ intermediate state of interest here, this decay may also 
receive important contributions from $NN^*$, $N^*N^*$ and $\Delta\Delta$, as well as 
other two-baryon and $NNm$ states; this may complicate the comparison with experiment 
considerably. Ref.\cite{Eaton:1983kb}, which has the largest event sample, finds 
a large but not dominant $\Delta\Delta$ contribution, 
$B(J/\psi \to \Delta^{++}{\bar \Delta}^{--}) = (1.10 \pm 0.29 ) \cdot 10^{-3}$, 
and gives a rather tight upper limit 
of $\approx 5\%$ on the contributing subprocess $J/\psi \to \pp\rho^0$;
$B(J/\psi \to \pp\rho^0) < 3.1 \cdot 10^{-4}$ (90\% c.l.).
As Ref.\cite{Eaton:1983kb} shows in their Fig.31 that the $\pi^+\pi^-$ 
invariant mass distribution from 
non-$\Delta\Delta$ events is a broad sigma-like distribution,
there may well be a large $J/\psi \to \pp f_0(\sim 500) \to \pp \pi^+\pi^-$ 
contribution to this decay, with a branching fraction comparable to the theoretical
$3.4\cdot 10^{-3}$ predicted for $g_{NNf_0} = 10$ (see Table~\ref{table:ppf0_rates}).
It will be very interesting to investigate this possible light $f_0$ contribution 
in a future experimental study, as well as to search for the 
$f_0(980)$ and $a_0(980)$ scalar states and the scalar glueball candidate $f_0(1500)$ in 
(higher-mass) charmonium decays, notably of the $\psi'$.

\subsection{Decays to $NN\omega$ and $NNV$}

Charmonium decays to $N\bar N \omega$ are especially interesting,
since the $\omega$ plays a crucial role in meson-exchange models of the $NN$ force,
as the dominant origin of the short-ranged ``hard core repulsion", through $t$-channel $\omega$ 
exchange. Conceptual problems with this $\omega$-exchange mechanism include 
1) the very small $NN$ separation implied by this mechanism 
($R_{NN}\approx 1/m_V \approx 0.3$~fm), at which quark-gluon dynamics may be a 
more appropriate description of the interaction, and 
2) the prediction of a corresponding short-ranged $N\bar N$ attraction and deeply bound $N\bar N$ states,
which are not observed.
(See for example Refs.\cite{Maltman:1983wx,Maltman:1983st,Barnes:1993nu}, and references cited therein.)

There are two strong coupling constants in the $\nnw$ vertex, as summarized by 
Eq.(\ref{eq:omega_vertex}),
the overall strength $g_{\nnw}$ of the Dirac ($\gamma_{\mu}$) coupling, 
and the relative strong magnetic Pauli coupling $\kappa_{\nnw}$ 
(here abbreviated $\gw$ and $\kw$, with $\aw = \gw^2/4\pi$). 
Unfortunately the $NNV$ couplings in the meson exchange models are not 
{\it a priori} well established 
experimentally, and are therefore usually fitted directly to $NN$ scattering data. 
These $NN$ scattering studies are thus fits to the data rather than predictions 
that test the assumed vector-meson-exchange scattering mechanism.
These $NN$ fits typically find $\gw \approx 10$-$15$ for the Dirac $NN\omega$ coupling,
whereas the $NN\omega$ Pauli coupling has remained poorly determined; 
examples of $NN\omega$ parameter sets from the $NN$ scattering literature include
$(\gw,\kw) = (12.2,-0.12)$~(Paris), 
$(12.5, +0.66)$~(Nijmegen), 
and 
$(15.9,0)$~(CD-Bonn) (these are cited in Ref.\cite{Downum:2006re}). 
Independent estimates of the $NN\omega$ coupling from experiment have been reported
by Sato and Lee \cite{Sato:1996gk} (from pion photoproduction) and by Mergell, 
Meissner and Drechsel \cite{Mergell:1995bf} (from nucleon EM form factors). 
Sato {\it et al.} assumed $\kw = 0$, and quoted the range $\gw = 7$-$10.5$ 
for experimentally favored values of the Dirac coupling. Mergell {\it et al.} 
found a small $\kw$ but a much larger Dirac coupling, 
$(\gw,\kw) = (20.86 \pm 0.25, -0.16 \pm 0.01)$.
Theoretical calculations include a QCD sum rule result of Zhu \cite{Zhu:1999kva}, who finds
$(\gw,\kw) = (18 \pm 8, 0.8 \pm 0.4)$, and a recent \3P0 quark model calculation of 
$NNm$ couplings \cite{Downum:2006re} which found the analytic result $\kw = -3/2$. 

Charmonium decays to $\ppw$ final states (and $N\bar NV$ more generally) may allow independent estimates
of these $NN\omega$ and $NNV$ couplings, again provided that they are dominated by the meson emission decay 
mechanism, or at least that this contribution to the decay amplitude can be isolated and quantified.
In the following discussion we will consider the decay $J/\psi \to \ppw$ as an example.

Results from experimental studies of the decay $J/\psi \to \ppw$ have been published by 
Mark-I \cite{Peruzzi:1977pb}, Mark-II \cite{Eaton:1983kb} and most recently 
BES-II \cite{Ablikim:2007dy}. The PDG value of the $J/\psi \to \ppw$ branching fraction, 
estimated from these results, is $B(J/\psi \to \ppw) = (1.10 \pm 0.15 )\cdot 10^{-3}$, which combined 
with the PDG $J/\psi$ total width gives an experimental partial width of
\be
\Gamma(J/\psi \to \ppw) = 103 \pm 14 \; \hbox{eV}.
\label{eq:Jpsi_ppw_width_pdg} 
\ee
The fact that this is approximately equal to the $\pppi$ partial width
(\ref{eq:Jpsi_pppi_width_pdg}) despite the much 
smaller phase space suggests a robust $\nnw$ coupling.

On evaluating this theoretical decay rate by integrating
Eq.(\ref{eq:Jpsi_ppw}) numerically with physical masses, both $NN\omega$ couplings free,
and (as used previously) $g_{J/\psi\pp}~=~1.62~\cdot~10^{-3}$, we find
\be
\Gamma(J/\psi \to \ppw) =  \aw \cdot \Big( 2.468  - 1.101 \,\kw + 0.886 \,\kw^2 \Big)\; \hbox{eV}.
\label{eq:Jpsi_ppw_width_thy} 
\ee

The single number $\Gamma(J/\psi \to \ppw)$ alone does not suffice to determine the 
$\nnw$ strong couplings since there are two free parameters, $\gw$ and $\kw$. 
If we set $\kw = 0$, following CD-Bonn \cite{Machleidt:2000ge} and
the Sato-Lee photoproduction study \cite{Sato:1996gk},
the measured partial width (\ref{eq:Jpsi_ppw_width_pdg}) 
and the theoretical decay rate (\ref{eq:Jpsi_ppw_width_thy})
imply $\gw = 23 \pm 3$ (provided that meson emission dominates this decay). 
This $\gw$ is rather larger than these references prefer for $\gw$; it is consistent 
however with the EM form factor fit of Mergell {\it et al.}~\cite{Mergell:1995bf}
and the range $18\pm 8$ reported by Zhu~\cite{Zhu:1999kva} from QCD sum rules.
If we instead assume the \3P0 quark model value $\kw = -3/2$ for the Pauli term 
\cite{Downum:2006re},
we find $\gw = 14.6 \pm 2.0$, which is consistent with the values quoted by 
$NN$ scattering studies, and is somewhat larger than the photoproduction value.

\begin{figure}[b]
\vskip -6mm
\includegraphics[width=0.9\linewidth]{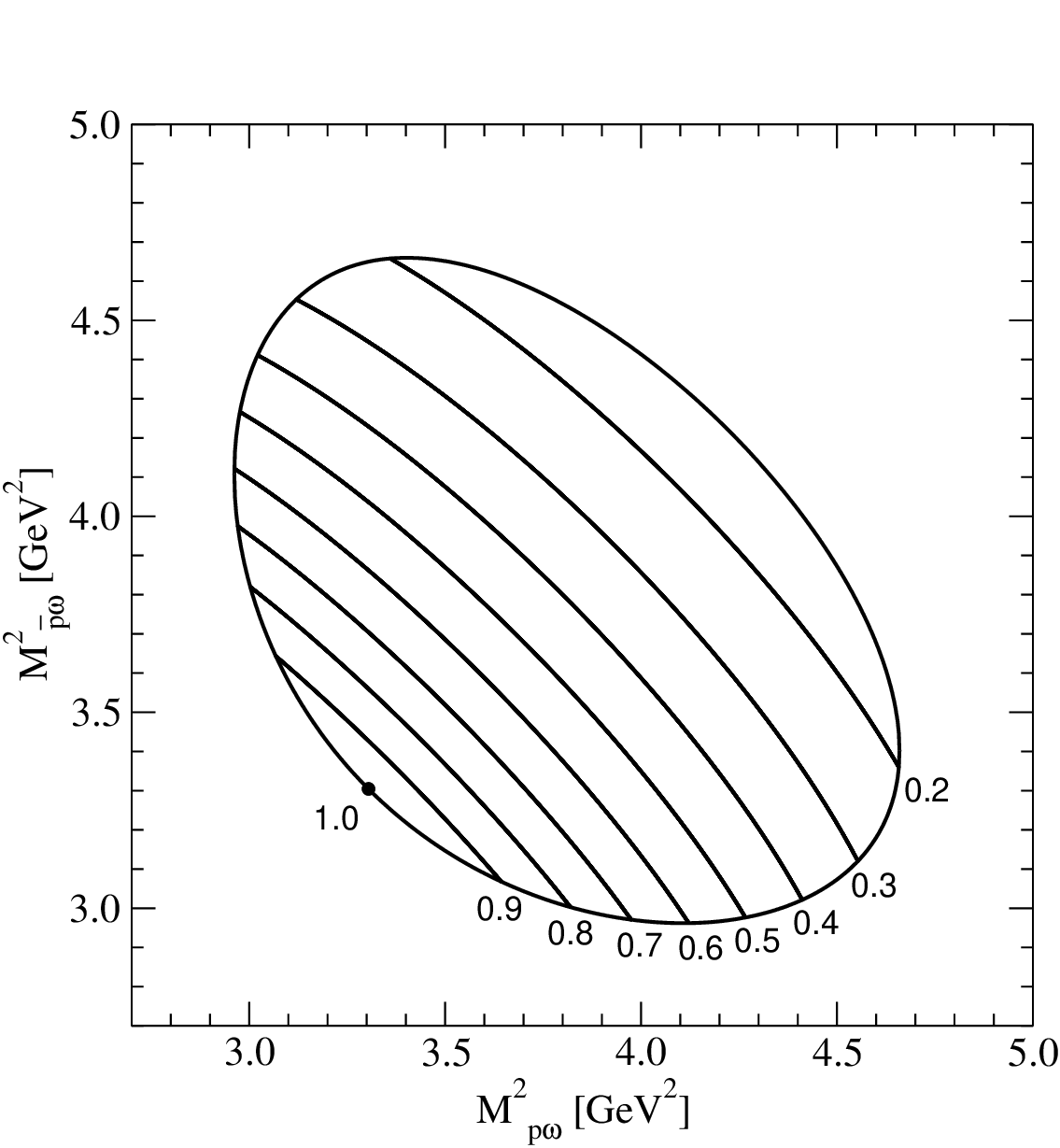}
\caption{Theoretical relative $J/\psi\to\ppw$ DP event density for $\kw = 0$, 
from Eq.(\ref{eq:Jpsi_ppw}). Note the suppression near $\pp$ threshold (upper right).}
\label{fig:blr_fig4}
\end{figure}
It is possible to estimate the two parameters $\gw$ and $\kw$ independently through a 
more detailed comparison between $J/\psi \to \ppw$ data and the theoretical DP event 
density, Eq.(\ref{eq:Jpsi_ppw}).
This theoretical event density is strongly dependent
on the Pauli coefficient $\kw$; with $\kw= 0$ the event density at lower $M_{\pp}$ is 
strongly suppressed (see Fig.\ref{fig:blr_fig4}). 

\begin{figure}[ht]
\vskip 6mm
\includegraphics[width=0.9\linewidth]{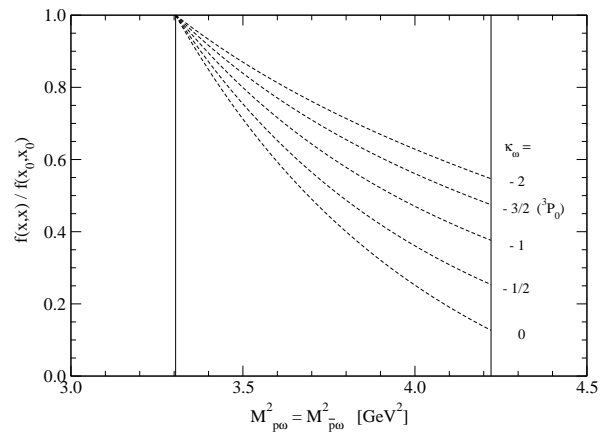}
\caption{Theoretical $J/\psi\to\ppw$ DP event density along the diagonal
$M^2_{p \omega} = M^2_{\bar p \omega}$, showing the strong $\kw$ dependence.}
\label{fig:blr_fig5}
\end{figure}

In contrast, for moderately large $|\kw |$, such as the
quark model value $\kw = -3/2$, the theoretical DP event density is closer to uniform.
This is illustrated in Fig.\ref{fig:blr_fig5}, which shows this event density along the diagonal
$M^2_{p \omega} = M^2_{\bar p \omega}$ (relative to the maximum value on diagonal) 
for various $\kw \leq 0$. 
If the meson emission model does give a good description of this decay, evidently it may be possible to 
determine $\kw$ by comparing the $J/\psi\to\ppw$ DP event density on diagonal to the prediction
in Fig.\ref{fig:blr_fig5}.

\subsection{Other $\Psi \to \pp V$ decays}

Other decays to $\pp V$ final states that are closely related to $J/\psi \to \ppw$ include
$J/\psi\to\pprho0$ and $J/\psi\to\ppphi$. In the meson emission model these are both described by the
decay formula (\ref{eq:Jpsi_ppw}), albeit with different vector meson masses and $NNV$ couplings.
Given the rounded PDG masses 
$m_{\rho^0} = 0.7755$~GeV 
and
$m_{\phi} = 1.0195$~GeV,
we predict the numerical decay widths 
\be
\Gamma(J/\psi \to \pprho0) =  \ar \cdot \Big( 2.614  - 1.155 \,\kr + 0.930 \,\kr^2 \Big)\; \hbox{eV}
\label{eq:Jpsi_pprho_width_thy} 
\ee
and
\be
\Gamma(J/\psi \to \ppphi) =  \aphi \cdot \Big( 0.184  - 0.109 \,\kphi + 0.087 \,\kphi^2 \Big)\; \hbox{eV}.
\label{eq:Jpsi_ppphi_width_thy} 
\ee

The $\rho^0$ case is especially interesting due to the range of values reported for $\kr$, as discussed by
Brown and Machleidt~\cite{Brown:1994pq}. Although vector dominance predicts $\kr = 3.7$ ``weak $\rho$", 
and some data has been interpreted as supporting this, 
fits to $\pi\pi \to N\bar N$ and S-D mixing in $NN$ scattering prefer a larger value ``strong $\rho$";
the Bonn potential model for example uses $(\ar,\kr) = (0.84, 6.1)$~\cite{Machleidt:1987hj}.  
A QCD sum rule calculation by Zhu~\cite{Zhu:1998yga} finds $(\gr,\kr) = (2.5\pm 0.2, 8.0\pm 2.0)$, 
comparable to the fitted Bonn values. In contrast, the valence quark model with a \3P0 $NN\rho$ coupling 
predicts a much smaller $\kr = -\kw = +3/2$~\cite{Downum:2006re}.
 
Using Eq.\ref{eq:Jpsi_pprho_width_thy} we can give meson emission model predictions for 
$\Gamma(J/\psi \to \pprho0)$ that follow from these various $(\gr,\kr)$ parameters. 
The Bonn parameters cited above give
$\Gamma(J/\psi \to \pprho0) = 25~\hbox{eV}$
and $B(J/\psi \to \pprho0) = 2.7\cdot 10^{-4}$; this is essentially equal to the current 
experimental upper limit~\cite{PDG,Eaton:1983kb} 
of $3.1\cdot 10^{-4}\ (90\%\; c.l.)$, which is a Mark-II result dating from 1984.
The Zhu QCD sum rule central values for $(\gr,\kr)$ give essentially identical results. 
In contrast the valence quark model with \3P0 
couplings $\gr = \gw /3$ and $\kr = +3/2$ (and using $\gr = 14.6$ from the $J/\psi \to \ppw$ discussion 
above) gives a much lower $\Gamma(J/\psi \to \pprho0) = 5.6~\hbox{eV}$
and hence $B(J/\psi \to \pprho0) = 6.0\cdot 10^{-5}$, which is about a factor of 5 below the 
current experimental limit.
The proximity of the Bonn and QCD sum rule parameter predictions for $B(J/\psi \to \pprho0)$ to the current 
limit suggests that an experimental study with significantly improved sensitivity could make 
a useful contribution to establishing $NN\rho$ couplings. 

The decay $J/\psi\to\ppphi$ in contrast {\it has} been observed, and has an experimental 
(PDG) branching fraction of
$B^{expt.}(J/\psi \to \ppphi) = (4.5\pm 1.5) \cdot 10^{-5}$, corresponding to
$\Gamma^{expt.}(J/\psi \to \ppphi) = 4.2 \pm 1.4 \; \hbox{eV}$. 
Unfortunately in this case we do not have a theoretical estimate for either $\aphi$ or $\kphi$, 
since the (valence level, leading-order) \3P0 model predicts no $NN\phi$ coupling.
Clearly it would be very interesting to obtain experimental values for these couplings, since 
little is known about the 
properties of Zweig-suppressed vertices. Again, if the meson emission model gives a good description of this decay,
a comparison of Eq.(\ref{eq:Jpsi_ppw}) to the experimental $J/\psi \to \ppphi$ DP event distribution 
should allow an experimental determination of the $NN\phi$ couplings.

Finally, we note in passing that $\psi'$ decays to $NNV$ are apparently not in agreement with 
the meson emission model; proceeding as above, we would predict a branching fraction of 
$B(\psi'\to \ppw) = 9.4 \cdot 10^{-4}$, whereas the PDG experimental value is an
order of magnitude smaller, $B^{expt.}(\psi'\to \ppw) = (6.9\pm 2.1) \cdot 10^{-5}$. 
Possible explanations for this discrepancy, including form factors and (destructive interference with) 
additional decay mechanisms, are discussed in the next section.
Since the total $\psi'\to \ppw$ data sample reported by CLEO \cite{Briere:2005rc} and BES \cite{Bai:2002yn}
comprises only about 35 events, it is not yet possible to establish 
the reason for this large discrepancy between experiment and the meson emission model. 
This would ideally involve a comparison between the predicted and observed Dalitz plot event densities. 
Hopefully this comparison will be possible using the large $\psi'$ data set being accumulated at BES-III. 

\section{Summary, Conclusions, and Future Developments}

In this paper we have presented and developed a hadron-level ``meson emission model" of 
charmonium decays of the type $\Psi \to \NNm$, where $\Psi$ is a generic charmonium resonance, 
$N$ is a nucleon and $m$ is a light meson. The model assumes that the decays take place 
through meson emission from the nucleon or antinucleon line, as a hadronic ``final state radiation" 
correction to a $\Psi \to \NN$ transition. As the model is relatively simple, we are able to evaluate 
the predicted DP event densities for many experimentally accessible and measured processes; 
in particular we give event densities for $\Psi = \eta_c$, $J/\psi$ (and $\psi'$), $\chi_{c0}$, 
$\chi_{c1}$ and $\psi'$ and $m = \pi^0, f_0$ and $\omega$, and implicitly all cases with the same 
$J^{PC}$ quantum numbers.

We used the reaction $J/\psi \to \pppi$ as a test case with no free parameters (the $J/\psi \pp$ and 
$NN\pi$ couplings are known), and compared the meson emission model predictions for the 
projected event density in $M_{p\pi}$ and the partial width $\Gamma(J/\psi \to \pppi)$ to experiment; 
the partial width is in good agreement, and the $M_{p\pi}$ event density appears to describe the 
non-$N^*$ ``background" contribution to this reaction observed experimentally. We also give 
predictions for $\Gamma(\Psi \to \pppi)$ for all $\Psi$ cases considered here; the overall trend of 
large and small widths and their approximate scale is reproduced by the model.

We also considered scalar and vector meson production. We estimated $\Psi \to \pp f_0$ branching fractions
for a light $f_0(500)$, and noted that the $J/\psi$ and $\psi'$ are favored for these studies, and the 
$\psi'$ is favored for glueball searches. In vector production we considered $J/\psi \to \ppw$ in particular, 
and noted that a high statistics study of this reaction could be used
to estimate the $NN\omega$ couplings ($\gw$ and $\kw$), which play a crucial role in meson exchange models
of the $NN$ force. We showed that the $J/\psi \to \ppw$ Dalitz plot event density
is rather sensitive to the poorly known $NN\omega$ Pauli coupling $\kw$. Determination of 
meson-nucleon strong couplings is a potentially very interesting application of $\Psi \to \ppm$ decays.

There are several theoretical developments that will be very important for future applications of this 
model. One should incorporate $N^*$ resonances; this is conceptually straightforward 
but may be technically complicated, as it will introduce many new and poorly known resonance coupling 
parameters and phases. This development is of course crucial to describe the data in reactions such as 
$J/\psi \to \pppi$, which clearly shows $N^*$ resonance peaks (Fig.\ref{fig:blr_fig3}). 
Another important development is the substitution of plausible $\Psi N^{(*)} \bar N^{(*)}$ and 
$N^{(*)}Nm$ hadron vertex form factors for the assumed coupling constants; 
the difficulty here is that hadronic form factors are poorly known, 
and models such as \3P0 that predict form factors have not been adequately developed and tested. 
Another interesting generalization of the strong vertices assumed here would be to include a $J/\psi\pp$ 
Pauli coupling; as we noted previously \cite{Barnes:2007ub}, this can explain the observed $e^+e^- \to J/\psi \to \pp$ 
angular distribution. Finally, one should include other significant decay mechanisms, as they become apparent 
through high-statistics studies of experimental Dalitz plots. These additional mechanisms might include 
intermediate meson resonances $m'$ that couple strongly to $\NN$, as in $\Psi \to m'm \to \ppm$; 
if the $m'$ resonances lie in the physical region, they will give rise to characteristic $m'$ resonance bands in 
$M_{\pp}$ that could be identified and incorporated in a more complete decay model. 
    
\section{Acknowledgments}

We are happy to acknowledge useful communications with
R.Mitchell, K.Seth, M.Shepherd, E.S.Swanson and 
B.S.Zou regarding this research, and Shu-Min Li and Xiao-Yan Shen in particular
for contributing the BES data used in preparing Fig.\ref{fig:blr_fig3}.
We also gratefully acknowledge the support of the 
Department of Physics and Astronomy of the University of Tennessee, 
the Physics Division of Oak Ridge National Laboratory, 
and the Department of Physics, the College of Arts and Sciences, and
the Office of Research at Florida State University.
This research was sponsored in part by the Office of Nuclear Physics, U.S. Department of Energy. 

\vfill\eject

\end{document}